\documentclass[fleqn,twoside,twocolumn,nofootinbib]{revtex4} 
\usepackage[sec]{ujp} 
\begin{document}
\title[ALEXANDER POLYNOMIAL INVARIANTS OF TORUS KNOTS $T(n,3)$
]
{ALEXANDER POLYNOMIAL INVARIANTS OF TORUS KNOTS
\boldmath$T(n,3)$
AND CHEBYSHEV POLYNOMIALS}%
\author{A.M. GAVRILIK }
\affiliation{Bogolyubov Institute for Theoretical Physics, Nat. Acad. of Sci. of Ukraine}
\address{14b, Metrolohichna Str., Kyiv 03680, Ukraine}
\email{omgavr@bitp.kiev.ua, pavlyuk@bitp.kiev.ua}
\author{A.M. PAVLYUK}
\affiliation{Bogolyubov Institute for Theoretical Physics, Nat. Acad. of Sci. of Ukraine}
\address{14b, Metrolohichna Str., Kyiv 03680, Ukraine}
\email{pavlyuk@bitp.kiev.ua} \udk{539.12; 517.984}  \razd{\secix}

\setcounter{page}{680}%
\maketitle

\newcommand{\ed}{\end{document}}
\newcommand{\be}{\begin{equation}}
\newcommand{\ee}{\end{equation}}
\newcommand{\bc}{\begin{center}}
\newcommand{\ec}{\end{center}}
\newcommand{\ba}{\begin{array}}
\newcommand{\ea}{\end{array}}
\newcommand{\ve}{\varepsilon}
\newcommand{\wt}{\widetilde}
\newcommand{\cl}{\centerline}
\newcommand{\ts}{\textstyle}

\begin{abstract}
The explicit formula, which expresses the Alexander polynomials
$\Delta_{n,3}(t)$ of torus knots $T(n,3)$ as a sum of the
Alexander polynomials $\Delta_{k,2}(t)$ of torus knots $T(k,2)$,
is found. Using this result and those from our previous papers,
we express the Alexander polynomials $\Delta_{n,3}(t)$ through
Chebyshev polynomials. The latter result is extended to
general torus knots $T(n,l)$ with $n$ and $l$ coprime.
\end{abstract}

\section{Introduction}
The interplay between knot theory and physics manifests itself
in various ways~\cite{At,Ka,Ka2,FV}.
Yet in 1975,  L.D. Faddeev proposed that knot-like solitons classified by the
integer-valued Hopf charge could be constructed in a modified sigma
model in the three-dimensional space \cite{Fa}. But only after
the paper by Faddeev and Niemi \cite{FN} in 1997,
the true hunting for knot configurations in theoretical physics
began~\cite{BS,GH, HS}.
Basing on the heavy usage of computer facilities, the conjecture
that the solitons with minimal energy might take the form of knots was
confirmed. The further increase of the computer power demonstrated that a
number of linked and knotted configurations, which are local or
global solutions with minimal energy, do exist in the well-known
Faddeev--Skyrme field-theory model.
  This issue was thoroughly elaborated in subsequent papers (e.g., in \cite{Su,RV})
  and is still developed.
    Let us recall that the concept and the role of knots is of
    basic importance for the non-perturbative sector of non-Abelian
  gauge theory (see~\cite{Kh} and references therein).

   As for the direct connection of knots with particle physics, one should
   mention the old paper by H. Jehle with the heuristic assigning of knot structures
   to real particles such as hadrons, neutrinos, and electrons \cite{Jeh}.
   More recently and on different grounds (using quantum groups or algebras)
   the connection "knots$ \leftrightarrow\!$ particles" was exploited in
   ~\cite{Ga1,Fi}. In particular, some set of torus knots appears
   in the context of meson phenomenology, namely in connection with
   quarkonia mass relations ~\cite{Ga1}.

It is worth to underline the importance (and, thus, a wide usage) of
applying different polynomial invariants of knots to modern physics
problems. E. Witten, for example, has shown that the Jones
polynomial invariants can be realized by the tools of topological
quantum field theory in $2+1$ dimensions \cite{Wi}.
 In addition, the Alexander polynomials and their well-known generalization,
 the HOMFLY polynomials, play a great role as well.

In our preceding paper~\cite{GP}, we studied the Alexander polynomial
invariants of torus knots $T(n,2)$ from the viewpoint of their
connection with Chebyshev polynomials, using the so-called
$q$-numbers.
  Therein, we explored a direct implication of the properties
of $q$-numbers and their generalization, $p,q$-numbers,
along with Chebyshev polynomials, for reproducing (from recursion
relations) the famous skein relations that determine the Alexander
polynomials and also HOMFLY polynomials.
We note that the relation of some knots to the Chebyshev
polynomials (in the context different from ours) was studied in~\cite{KP}.
It should be emphasized
that our whole treatment in paper~\cite{GP} concerned with the
simplest, though nontrivial, set of torus knots -- the series
$T(n,2)$, $n$ being an odd integer.
Since the Alexander (and HOMFLY) polynomials provide the most
important and widely used characteristics of knots and links, their
thorough investigation is believed to be helpful for the further
understanding of the properties of knot-like structures from the first
principles and for clarifying the physical interpretation of these
knot invariants.

 In the present paper, we first concentrate on the study of the Alexander polynomial
 invariants $\Delta_{n,3}(t)$ for the set of torus knots $T(n,3)$, with
 $n$ and $3$ coprime, and on their close connection with relevant Chebyshev
 polynomials.
   Then, we extend our treatment to the general class of torus knots
   $T(n,l)$,  $l\ge 2$,
   with $n$ and $l$  coprime.
   Like in our preceding paper, the concept of $q$-numbers turns out
   to be very helpful.

\section{Chebyshev Polynomials}

Since we will exploit the Chebyshev polynomials below, let us give some
sketch of them.
  Chebyshev polynomials of the first kind $T_{n}(x)$ can be defined as
  \be\label{T} T_{n}(x)=2\cos
(n\theta)\,,\qquad 2\cos\theta=x\,.
  \ee
  This normalization means that the Chebyshev polynomials are monic
  (with unit coefficient at
$x^{n}$).
These Chebyshev polynomials, with account of the formula
\[
\cos{(n+1)\theta}+\cos{(n-1)\theta}=2\cos{\theta}\cos{n\theta}
 \]
are seen to satisfy
 the relation
 \be\label{rec-T} T_{n+1}=xT_{n}-T_{n-1}\,,\qquad T_{0}=2\,,\quad
T_{1}=x\,, \ee
 which is nothing but the recurrence relation.
The latter can be used as an alternative definition of $T_{n}(x)\,.$
Some few low-degree cases of $T_{n}(x)$ read
 \[ \label{Tx}
 \ba{l}
T_{0}=2\,,\quad T_{1}=x\,,\quad
   T_{2}=x^{2}-2\,,\quad
T_{3}=x^{3}-3x\,,
\vspace{2mm}  \\
T_{4}=x^{4}-4x^{2}+2\,,\quad
T_{5}=x^{5}-5x^{3}+5x\,.
 \ea \]

Chebyshev polynomials of the second kind can be defined as
  \be\label{V} V_{n}(x)={\sin(n+1)\theta\over {\sin\theta}}\,,\qquad
2\cos\theta=x
  \ee
 or through the recurrence relation
 \be\label{rec-V} V_{n+1}=xV_{n}-V_{n-1}\,,\qquad V_{0}=1\,,\quad
V_{1}=x\,.
 \ee
  First few low-order Chebyshev polynomials of
the second kind are as follows:
 \[
 \ba{l}
V_{0}=1\,,\quad V_{1}=x\,,\quad V_{2}=x^{2}-1\,,\quad
V_{3}=x^{3}-2x\,,
\vspace{2mm}  \\
V_{4}=x^{4}-3x^{2}+1\,,\quad V_{5}=x^{5}-4x^{3}+3x\,.
 \ea \]
There exists a connection between $T_{n}(x)$ and $V_{n}(x)$, namely
 \be\label{con1}
 T_{n}(x)=V_{n}(x)-V_{n-2}(x)\, , \quad\quad \quad  n\ge 1\,,\ee
where, and throughout the text, we omit polynomials with negative
indices (at $n=1\,$ we omit $V_{-1}(x)$).

Note that Chebyshev polynomials of the first kind can be presented
in the form
  \be\label{T-t} T_{n}(x)=t^{n}+t^{-n}\,, \qquad
t+t^{-1}=x\,,
 \ee
 while Chebyshev polynomials of the second kind
as
  \be\label{V-t}
V_{n}(x)={{\textstyle{t^{n+1}-t^{-n-1}}}\over\textstyle{t-t^{-1}}}\,,
\qquad t+t^{-1}=x\,.
 \ee
 This is obvious if we put
$t=e^{i\theta}\,.$

\section{Alexander Polynomials for Torus Knots }

 The Alexander polynomials $\Delta(t)$ for knots and links can be defined
by the skein relation~\cite{Ro}
 \be\label{alex-skein}
\Delta_{+}(t)=(t^{1\over2}-t^{-{1\over2}})\Delta_{O}(t)+\Delta_{-}(t)\,
 \ee
 along with the normalization condition (for an unknot)
  \be\label{unknot} \Delta_{\mathrm{unknot}}=1\,.\ee

 Using (\ref{alex-skein})-(\ref{unknot}), one can find the Alexander
 polynomial for any knot or link by applying, in a standard way, the
 operations of ``switching'' and ``elimination''.

We focus on the torus knots $T(n,l),$ where $n$ and $l$ are coprime
positive integers (the torus knots $T(n,l)$ and $T(l,n)$ are
equivalent).
   As known \cite{Ro}, the Alexander polynomial invariant for  
$T(n,l)$ is given by the formula
  \be\label{delta}
\wt\Delta_{n,l}(t)={{(t^{nl}-1)(t-1)}\over{(t^{n}-1)(t^{l}-1)}}\,.
  \ee
For $l=2\,$
 this yields
\be\label{delta2}
\wt\Delta_{n,2}(t)={{t^n+1}\over{t+1}}\,,\qquad n=1, 3, 5, 7, ...\,.
\ee
 Let us list  few examples of polynomial (\ref{delta2}):
\[
\wt\Delta_{1,2}(t)=1\,,\quad \wt\Delta_{3,2}(t)=t^{2}-t+1\,,\quad
\]
\[
\wt\Delta_{5,2}(t)=t^{4}-t^{3}+t^{2}-t+1\,.
\]
For $l=3\,,$  relation (\ref{delta}) yields
 \be\label{delta3}
\wt\Delta_{n,3}(t)={{t^{2n}+t^n+1}\over{t^{2}+t+1}}\,, \qquad n=1,
2, 4, 5, 7, 8, ...\,.
 \ee
The first entries of (\ref{delta3}) are
\[
\wt\Delta_{1,3}(t)=1\,,\quad \wt\Delta_{2,3}(t)=t^{2}-t+1\,,
\]
\[
\wt\Delta_{4,3}(t)=t^{6}-t^{5}+t^{3}-t+1\,.
 \]
 Note that $\wt\Delta_{1,2}(t)$, $\wt\Delta_{1,3}(t),$ and $\wt\Delta_{1,l}(t)$
 all agree with (\ref{unknot}). The same concerns $\wt\Delta_{n,1}(t)$, see (\ref{delta}).

The Alexander polynomials (\ref{delta}) are reduced to the standard
polynomial form and have the degree $nl-n-l+1=(n-1)(l-1)$.
 However, the Alexander polynomials $\wt\Delta_{n,l}(t)$ can
 also be presented in the form of the Laurent polynomials ${\Delta}_{n,l}(t)\,$
possessing the explicit $t\leftrightarrow t^{-1}$ symmetry.
 This (Laurent type) form, with account of the multiplier
   $t^{-{{nl-n-l+1}\over {2}}}$,
   satisfies the relation (note the absence of a tilde on the l.h.s.)
\be\label{deltal} {\Delta}_{n,l}(t)=\wt\Delta_{n,l}(t)\cdot {\ts
t^{-{{nl-n-l+1}\over {2}}}}={{(\ts t^{nl\over 2}-t^{-{nl\over
2}})(t^{1\over 2}-t^{-{1\over 2}})} \over{(\ts t^{n\over
2}{-}t^{-{n\over 2}})(t^{l\over 2}{-}t^{-{l\over 2}})}}\,.
 \ee
  By $m,$ we denote the highest positive degree of the Alexander
  polynomial $\Delta_{n,l}(t)\,$ given as a Laurent polynomial in (\ref{deltal}):
\be\label{max}m={1\over2}(n-1)(l-1)\,.\ee
  From now on, we consider two cases:
\be\label{23} l=2\,,\quad n=2m+1\,;\qquad l=3\,,\quad n=m+1\,.
\ee
For $l=2\,,$ Eq. (\ref{deltal}) gives
 \be\label{deltal2}
{\Delta}_{n,2}(t)= {{t^{n\over 2}+t^{-{n\over 2}}} \over{t^{1\over
2}+t^{-{1\over 2}}}}\,
 \ee
  of which the first several examples ($n{=}1, 3, 5$) are
\[
{\Delta}_{1,2}(t)=1\,,\quad \ \ \  {\Delta}_{3,2}(t)=t-1+t^{-1}\,,
\]
\[
{\Delta}_{5,2}(t)=t^{2}-t+1-t^{-1}+t^{-2}\,.\]
If $l=3\,,$ (\ref{deltal}) yields 
 \be\label{deltal3} {\Delta}_{n,3}(t)= {{t^{n}+1+t^{-{n}}}
\over{t+1+t^{-{1}}}}\,.
 \ee
 A few examples for $n{=}1, 2, 4$ are
\[
{\Delta}_{1,3}(t)=1\,,\quad \ \ \ \ {\Delta}_{2,3}(t)=t-1+t^{-1}\,,
\]
\[
{\Delta}_{4,3}(t)=t^{3}-t^{2}+1-t^{-2}+t^{-3}\]
(note the coincidence of $\Delta_{2,3}(t)$ with $\Delta_{3,2}(t)$
above).

 Below, we find a connection between the Alexander
polynomials for $T(n,3)$ and $T(k,2)\,$ torus knots.

\section{Presenting \boldmath$\Delta_{n,3}(t)$ in Terms of \boldmath$\Delta_{k, 2}(t)$}

The goal of this section is to express any Alexander polynomial
$\Delta_{n, 3}(t)$ as an algebraic sum of Alexander polynomials for
$T(k,2)$ torus knots.

{\bf{Proposition 1.}}
 {\it The (Laurent form) Alexander
polynomials for the torus knots $T(n,3)$ and the torus knots
$T(k,2)$ are connected through the relation}
\be\label{t1}
{\Delta}_{n,3}(t)-{\Delta}_{n-3,3}(t)={\Delta}_{2n-1,2}(t)-{\Delta}_{2n-5,2}(t)\,.
\ee
The proof goes by straightforward checking with the use of formulas
(\ref{deltal2}) and (\ref{deltal3}).

{\bf{Proposition 2.}} 
{\it The Alexander polynomial of torus knot $\, T(n,3)$ is
expressible as a sum of the Alexander polynomials of torus knots
$\, T(k,2)$ by the formula\footnote{ Note that the remark similar to
the one after (5) is applied here too.}
 \be\label{t2} {\Delta}_{n,3}(t)
=\sum\limits_{j=0}^{d}\bigl({\Delta}_{2n-1-6j,\, 2}(t)-
{\Delta}_{2n-5-6j,\, 2}(t)\bigr)\,,
  \ee
where $d$ is the integer part of $\frac{2n-1}{6}$:
$\,\,\,d=\Bigl[\frac{2n-1}{6}\Bigr]$}.

To prove (\ref{t2}), we rewrite Eq. (\ref{t1}) first as
 \be\label{t1a}
{\Delta}_{n,3}(t)={\Delta}_{2n-1,2}(t)-{\Delta}_{2n-5,2}(t)+{\Delta}_{n-3,3}(t)\,
 \ee
and, shifting $n$ to $n-3\,,$ we deduce the relation
\be\label{t1b}
{\Delta}_{n-3,3}(t)={\Delta}_{2n-7,2}(t)-{\Delta}_{2n-11,2}(t)+{\Delta}_{n-6,3}(t)\,\ee
from (\ref{t1a}). Substituting
(\ref{t1b}) in (\ref{t1a}) yields
\[
{\Delta}_{n,3}(t)={\Delta}_{2n-1,2}(t)-{\Delta}_{2n-5,2}(t)
+{\Delta}_{2n-7,2}(t)-
\]
\be\label{t1c}-{\Delta}_{2n-11,2}(t)+{\Delta}_{n-6,3}(t)\,.\ee
Similarly to (\ref{t1b}), we further expand ${\Delta}_{n-6,3}(t)$
and put the result into (\ref{t1c}),
and so on.
  At the last stage of the process for ${\Delta}_{n,3}(t),$
  we come to ${\Delta}_{2,3}(t)$ and/or ${\Delta}_{1,3}(t).$
 Since
${\Delta}_{2,3}(t)={\Delta}_{3,2}(t)$ and ${\Delta}_{1,3}(t)
=1={\Delta}_{1,2}(t)\,,$ we obtain the expression for the Alexander
polynomial of torus knot $T(n,3)$ given through a sum of the
Alexander polynomials for torus knots $T(n,2)$ only.

It is worth to give, from (\ref{t2}), a few examples of expressing the
Alexander polynomials ${\Delta}_{n,3}(t)$ by a sum of the
Alexander polynomials ${\Delta}_{n,2}(t):$
 \[
{\Delta}_{1,3}(t)={\Delta}_{1,2}(t)\,,
\]
\[
{\Delta}_{2,3}(t)={\Delta}_{3,2}(t)\,
\]
\[
{\Delta}_{4,3}(t)={\Delta}_{7,2}(t)-{\Delta}_{3,2}(t)+{\Delta}_{1,2}(t)\,,
\]
\[
{\Delta}_{5,3}(t)={\Delta}_{9,2}(t)-{\Delta}_{5,2}(t)+{\Delta}_{3,2}(t)\,,
\]
\[
{\Delta}_{7,3}(t)={\Delta}_{13,2}(t)-{\Delta}_{9,2}(t)+{\Delta}_{7,2}(t)-
\]
\[
-{\Delta}_{3,2}(t)+{\Delta}_{1,2}(t)\,,
\]
\[
{\Delta}_{8,3}(t)={\Delta}_{15,2}(t)-{\Delta}_{11,2}(t)+{\Delta}_{9,2}(t)-
\]
\[
-{\Delta}_{5,2}(t)+{\Delta}_{3,2}(t)\,,
\]
\[
{\Delta}_{10,3}(t)={\Delta}_{19,2}(t)-{\Delta}_{15,2}(t)+{\Delta}_{13,2}(t)-
\]
 \be \label{exam32}
-{\Delta}_{9,2}(t)+{\Delta}_{7,2}(t)-{\Delta}_{3,2}(t)+{\Delta}_{1,2}(t)\,.
\ee
 Thus, we have obtained the formula expressing (each from) the set
 ${\Delta}_{n,3}(t)$ through the sum of a definite number of the
 Alexander polynomials ${\Delta}_{k,2}(t)$ with proper signs.

It would be of interest to generalize this result to any torus knot
${\Delta}_{n,l}(t)$.

\section{Expressing the Alexander Polynomials by Chebyshev Polynomials}

In this section, we present the Alexander polynomials
$\Delta_{n,2}(t)$ and $\Delta_{n,3}(t)$ in terms of the Chebyshev
polynomials $T_{n}(x)$ and $V_{n}(x)\,,\,\, x=t+t^{-1}\,. $

{\bf{Proposition 3.}} 
{\it There exists the following relation between the Alexander
polynomials $\Delta_{n,2}(t)$
for torus knots $T(n,2)$ and Chebyshev polynomials of the second
kind:}
\[
 \Delta_{n,2}(t)\equiv\Delta_{2m+1,2}(t)=
\]
\be\label{gavr}
 =V_{m}(x)-V_{m-1}(x)\,,\quad
x=t+t^{-1}\,.\ee

To prove (\ref{gavr}), we use formula (\ref{deltal2}), along
with $V_{m-1}(t)$ and $V_{m}(t)\,$ taken from (\ref{V-t}).
Note that an analog of (\ref{gavr}) was first found in \cite{Ga1} on
the base of using the $q$-numbers.

Equation (\ref{gavr}) can also be written as
\be\label{gavr2} \Delta_{n,2}(t)
=V_{\frac{n-1}{2}}(x)-V_{\frac{n-3}{2}}(x)\,,\quad x=t+t^{-1}\,.\ee
From (\ref{gavr}) and (\ref{con1}), we find ($m\ge1$):
 \be\label{tdd}
T_{m}(x)=\Delta_{2m+1,2}(t)+\Delta_{2m-1,2}(t)\,,\quad
t+t^{-1}=x\,.
 \ee
  Hence, Eq. (\ref{tdd}) expresses the Alexander
polynomials in terms of Chebyshev polynomials of the first kind.
 Here are some examples:
\[
\Delta_{3,2}(t)=T_{1}(x)-\Delta_{1,2}(t)=T_{1}(x)-1\,, 
\]
\[
\Delta_{5,2}(t)=T_{2}(x)-T_{1}(x)+1\,,
\]
\be
 \label{dt}
\Delta_{7,2}(t)=T_{3}(x)-T_{2}(x)+T_{1}(x)-1\,.
\ee
 The general formula
covering (\ref{dt}) is
\[
\Delta_{2m+1,2}(t){=}T_{m}(x)-T_{m-1}(x)+T_{m-2}(x){-}\cdots{+}
\]
 \be \label{dt-gen}
+(-1)^{m-1}T_{1}(x) +(-1)^{m}
=\sum\limits_{k=0}^{m-1}(-1)^{k}T_{m-k}(x) +(-1)^{m} \,. \ee

{\bf{Proposition 4.}} 
{\it The Alexander polynomial for torus knot $T(n,3)$ is expressed
as the sum of Chebyshev polynomials of the second kind by the
relation}
\[
  {\Delta}_{n,3}(t)
=V_{n-1}(x)+
\]
 \be \label{exam32}
+\sum\limits_{k=0}^{d}\bigl(-V_{n-2-3k}(x)-V_{n-3-3k}(x)+
2V_{n-4-3k}(x)\bigr),
 \ee
{\it  where} $d=\Bigl[\frac{n-2}{3}\Bigr],$  $x=t+t^{-1}.$

This immediately follows from (\ref{t2}) and (\ref{gavr}).

From formula (\ref{exam32}), we have
\[
 \Delta_{n,3}(t)=V_{n-1}(x)-V_{n-2}(x)-V_{n-3}(x)+
\]
\[
+2V_{n-4}(x)-
V_{n-5}(x)-V_{n-6}(x)+
\]
\be\label{p4a} +2V_{n-7}(x)-V_{n-8}(x)-\cdots,\quad
x{=}t+t^{-1}\,.\ee

Some first examples of (\ref{p4a}) are
\[
\Delta_{1,3}(t)=V_{0}(x),\quad \Delta_{2,3}(t)=V_{1}(x)-V_{0}(x),
\]
\[
\Delta_{4,3}(t)=V_{3}(x)-V_{2}(x)-V_{1}(x)+2V_{0}(x),
\]
\[
\Delta_{5,3}(t)=V_{4}(x)-V_{3}(x)-V_{2}(x)+2V_{1}(x){-}V_{0}(x),
\]
\[
\Delta_{7,3}(t)=V_{6}(x)-V_{5}(x)-V_{4}(x)+2V_{3}(x){-}V_{2}(x){-}
\]
\[
\label{p4b} -V_{1}(x)+2V_{0}(x) .
\]

{\bf Proposition 5.} 
{\it The Alexander polynomial for the torus knot $T(n,3)$ is
expressed through Chebyshev polynomials of the first kind by the
formula}
 \be\label{p5} {\Delta}_{n,3}(t)
{=}\sum\limits_{k=0}^{d}\bigl(T_{n-1-3k}(x)-T_{n-2-3k}(x)\bigr){+}(-1)^{n-d},
 \ee
{\it  where } $ d=\Bigl[\frac{n-1}{3}\Bigr],$   $x=t+t^{-1}. $

The proof follows from the joint use of (\ref{t2}) and (\ref{tdd}).

Here are some special cases of (\ref{p5}):
  \[
\label{p5a} \Delta_{1,3}(t)=T_{0}(x)-1,\quad
\Delta_{2,3}(t){=}T_{1}(x){-}T_{0}(x){+}1,
\]
\[
\Delta_{4,3}(t)=T_{3}(x)-T_{2}(x)+T_{0}(x)-1,
\]
\[
\Delta_{5,3}(t)=T_{4}(x)-T_{3}(x)+T_{1}(x)-T_{0}(x)+1,
\]
\[
\Delta_{7,3}(t)=T_{6}(x)-T_{5}(x)+T_{3}(x)-T_{2}(x){+}
\]
\[
+T_{0}(x)-1
 .\]

\section{Some Useful Formulas
         for the Alexander Polynomials}
Putting definition (\ref{V}) of the Chebyshev
polynomials in (\ref{gavr}), we obtain the following formula for the Alexander
polynomials $\Delta_{2m+1,2}(t):$
 \be\label{gavr-cos} \Delta_{2m+1,2}(t)=
{\cos(m+{1\over2})\theta\over\cos{\theta\over2}}\,\qquad
2\cos\theta=t+t^{-1}\,,\ee
or, setting  $m={1\over2}(n-1)\,,$
\be\label{gavr-cos2}
\Delta_{n,2}(t)=
{\cos{n\theta\over2}\over\cos{\theta\over2}}\,,\qquad
2\cos\theta=t+t^{-1}\,,\ee where $n=1, 3, 5, 7, \cdots \,.$
From the latter, remembering that  
 $\cos\theta=2\cos^{2}{\theta\over2}-1\,,$ we have
 \be\label{AT1} \Bigl(\Delta_{n,2}(t)\Bigr)^{2}=
{{T_{n}(x)+2}\over{x+2}}\,,\qquad x=t+t^{-1}\,.
 \ee
From (\ref{gavr-cos2}) and (\ref{V}), we find
\be\label{AVV} \Delta_{n,2}(t)=
{{V_{2n-1}(y)}\over{yV_{n-1}(y)}}\,,\qquad y^{2}=t+t^{-1}+2\,. \ee

Rewriting (\ref{T-t}) as
 \be\label{T-t2}
T_{n}(y)=t^{n\over2}+t^{-{n\over2}}\,, \qquad
y=t^{1\over2}+t^{-{1\over2}}\, \ee
and putting it in
(\ref{deltal2}), we have
  \be\label{AT2-} \Delta_{n,2}(t)=
{{T_{n}(y)}\over{y}}\,.
 \ee

Using (\ref{deltal3}) and (\ref{T-t2}), we obtain the formula for
$\Delta_{n,3}(t)$ in terms of $T_{n}(y)$:
 \[  
\Delta_{n,3}(t)=
{\textstyle{{T_{n}^{2}(y)-1}}\over{\textstyle{y^{2}-1}}}\, ,
 \]
 where $n=1, 2, 4, 5, 7, ...\,.$
Using (\ref{T-t}), Eq. (\ref{deltal3}) can also be written in the
form
 \[ 
  \Delta_{n,3}(t)=
{\textstyle{{T_{n}(x)+1}}\over{\textstyle{x+1}}}\,,\qquad
x=t+t^{-1}\,.
 \]

\section{Alexander Polynomials of \boldmath$T(n,l)$ in Terms of \boldmath$q$-Numbers}

In this section, we extend the formula (\ref{gavr}) valid for
$l=2$ to the general case of arbitrary $l\,$.
 To this end, like in \cite{GP}, we use the $q$-numbers~\cite{Bi,Ma,Kach,Kac}.
 Recall that the $q$-number corresponding to an integer $n$
 is defined as
 \be \label{q-def} [n]_{q}={{\textstyle
{q^{n}-q^{-n}}}\over{\textstyle {q-q^{-1}}}}\,,
 \ee
  with $q$ being a parameter.
  If $q\rightarrow 1\,,$ we recover $n$:
$[n]_{q}\stackrel{q\to 1}{\longrightarrow}n$.
If $\,q\,$ is viewed as a variable, we arrive at $q$-polynomials.

With account of (\ref{q-def}), Eq. (\ref{deltal}) can be immediately
rewritten in terms of $q$-numbers to yield $(q\equiv t)$:
  \be\label{dnl-q}
\Delta_{n,l}(t)=\frac{[nl]_{t^{1\over2}}}{{[n]_{t^{1\over2}}}{[l]_{t^{1\over2}}}}=
\frac{[n]_{t^{l\over2}}}{{[n]_{t^{1\over2}}}}=
\frac{[l]_{t^{n\over2}}}{{[l]_{t^{1\over2}}}}
  \,.\ee

From (\ref{V-t}) and (\ref{q-def}), we have
 \be\label{V-q}
V_{n}(x)=
[n+1]_{q}\,, \qquad x=q+q^{-1}\,.
 \ee

Formula (\ref{gavr}) in terms of $q$-numbers reads
 \be\label{AV2q} \Delta_{2m+1,2}(t)=[m+1]_{t}-[m]_{t}\,,
 \hspace{10mm} t\equiv q \ ,
 \ee
or, since $n=2m+1$,
 \be\label{AV2qn}
\Delta_{n,2}(t)=\Bigl[{{n+1}\over2}\Bigr]_{t}-\Bigl[{{n-1}\over2}\Bigr]_{t}\,.\ee
It is easy to verify that (\ref{AV2qn}) is reduced to (\ref{deltal2}).
If $l=3$, relation (\ref{deltal3}) yields the formula somewhat similar
to (\ref{AV2qn}):
 \be\label{AV3qn}
\Delta_{n,3}(t)=\Bigl[{{2n+1}\over3}\Bigr]_{t^{3\over2}}-\Bigl[{{2n-1}\over3}\Bigr]_{t^{3\over2}}
+ \Bigl[{{1}\over3}\Bigr]_{t^{3\over2}} \,.
 \ee
 Each of the three terms can be rewritten, because
 \[
 \Bigl[{{X}\over3}\Bigr]_{t^{3\over2}}={[{X}]_{t^{1\over2}}\over
{[3]_{t^{1\over2}}}}\,.
 \]
As a result, Eq. (\ref{AV3qn}) becomes
 \be\label{AV3qn-}
\Delta_{n,3}(t)={1\over{[3]_{t^{1\over2}}}}
\Bigl([2n+1]_{t^{1\over2}}- [2n-1]_{t^{1\over2}} +{1}\Bigr) \,.
 \ee

In the case of arbitrary $l$, the generalization of (\ref{AV2qn}) and
(\ref{AV3qn})
  can be obtained from (\ref{deltal}), namely
 \be\label{deltall}\ba{l}
\Delta_{n,l}(t){=}{\Bigl[{{n(l-1)+1}\over l}\Bigr]_{t^{l\over2}}-
\Bigl[{{n(l-1)-1}\over l}\Bigr]_{t^{l\over2}}}+
\frac{[l-2]_{t^{\frac{n}{2}}}}{[l]_{t^{\frac{1}{2}}}}\,.\ea\ee
From this, we deduce a generalization of Eq. (\ref{AV3qn-}):
\[
\Delta_{n,l}(t)=\frac{[n(l-1)+1]_{t^{1\over2}}}{[l]_{t^{1\over2}}}-
\frac{[n(l-1)-1]_{t^{1\over2}}}{[l]_{t^{1\over2}}}+
\]
\be\label{deltall-}
+\frac{[n(l-2)]_{t^{1\over2}}}{[n]_{t^{1\over2}}}\cdot
\frac{1}{[l]_{t^{1\over2}}}  \, .
  \ee

\section{Alexander Polynomials of \boldmath$T(n,l)$
                      and Chebyshev Polynomials}

We can express (\ref{AV3qn-}) in terms of the Chebyshev polynomials:
  \be\label{AV3qn--}
\Delta_{n,3}(t)=\frac{V_{2n}(y)-V_{2n-2}(y)+1}{V_{2}(y)}, \qquad
y{=}t^{1\over2}+t^{-{1\over2}}.
 \ee
Likewise, as an extension of the latter to the  $T(n,l)$ torus
knots, relation (\ref{deltall-}) yields
\be\label{deltall--}\ba{l} \Delta_{n,l}(t)=\frac{\ts
V_{n(l-1)}(y)-V_{n(l-1)-2}(y)}{\ts V_{l-1}(y)}+
\vspace{2mm}\\
+\frac{\ts V_{n(l-2)-1}(y)}{\ts V_{l-1}(y)\cdot V_{n-1}(y)}\,,
\qquad y=t^{1\over2}+t^{-{1\over2}}. \ea
  \ee
  In addition, using Eqs. (\ref{dnl-q}), we obtain another three
formulas giving $\Delta_{n,l}(t)$ via the Chebyshev polynomials:
  \be\label{dnl-v1} \Delta_{n,l}(t)=\frac{V_{nl-1}(y)}
{V_{n-1}(y)\cdot V_{l-1}(y)}\,, \qquad
y=t^{1\over2}+t^{-{1\over2}}\,
 \ee
 or
 \be\label{dnl-v2}
\Delta_{n,l}(t)=\frac{V_{n-1}(z_{1})} {V_{n-1}(y)}, \quad
z_{1}{=}t^{l\over2}+t^{-{l\over2}}, \quad
y{=}t^{1\over2}+t^{-{1\over2}},
  \ee or
\be\label{dnl-v3} \Delta_{n,l}(t)=\frac{V_{l-1}(z_{2})}
{V_{l-1}(y)}, \quad z_{2}{=}t^{n\over2}+t^{-{n\over2}}, \quad
y{=}t^{1\over2}+t^{-{1\over2}}. \ee

Now let us pay attention to the following very interesting point:
it follows from (\ref{T-t2}) and (\ref{dnl-v2}) that
\be\label{dnl-v2-}
\Delta_{n,l}(t)=\frac{V_{n-1}\bigl(T_{l}(y)\bigr)} {V_{n-1}(y)},
\qquad y{=}t^{1\over2}+t^{-{1\over2}} .
 \ee
  Analogously, from (\ref{T-t2}) and (\ref{dnl-v3}), we obtain
  \be\label{dnl-v3-}
\Delta_{n,l}(t)=\frac{V_{l-1}\bigl(T_{n}(y)\bigr)} {V_{l-1}(y)},
\qquad y{=}t^{1\over2}+t^{-{1\over2}}.
 \ee
  As is seen, the numerators of both Eqs. (\ref{dnl-v2-}) and (\ref{dnl-v3-})
  include the expression $V(\ldots )$, i.e. a Chebyshev polynomial of the
second kind, in which the role of an argument, in turn, is played by the
Chebyshev polynomial of the first kind.

\section{Dependence of \boldmath$\Delta_{n,l}(t)$ on \boldmath$n$ Through
         \boldmath$\Delta_{n,2}(t)$ }

In this section, we show that the whole dependence of the Alexander
polynomial $\Delta_{n,l}(t)$ on the number $n$ can be given solely
 through some lower Alexander polynomial $\Delta_{n,k}(t)\,,\,\,k<l\,,$
 $k$ being a fixed positive integer coprime both with $l$ and $n$.
  Let us first consider the simplest case $k=2$.

Rewriting  (\ref{deltal2}) in the form
 \[\label{deltal2z}
\Delta_{n,2}(t)= \frac{ Z+Z^{-1} }{t^\frac{1}{2}+t^\frac{-1}{2}} ,
\hspace{12mm} Z\equiv t^{n\over2}\, ,
 \]
 we arrive at the equation
 \[
Z^{2}-(\ts t^{1\over2}+t^{-{1\over2}}) \Delta_{n,2}(t) Z+1=0
 \]
and at the identical one for $Z^{-1}$, which have the solutions
 \[
 Z
={1\over2}\Bigl(( t^{1\over2}+t^{-{1\over2}}) \Delta_{n,2}(t)
 \pm
\]
\[
\pm {\sqrt{{(\ts t^{1\over2}+t^{-{1\over2}})^{2}
(\Delta_{n,2}(t))^{2}-4}}}
 \Bigr)\, ,
   \]
\[
 Z^{-1} 
 ={1\over2}\Bigl(( t^{1\over2}+t^{-{1\over2}})
\Delta_{n,2}(t) \mp
\]
\[ \mp {\sqrt{{(\ts t^{1\over2}+t^{-{1\over2}})^{2}
(\Delta_{n,2}(t))^{2}-4}}}\,\,
 \Bigr)\,.
   \]
 This allows us to present
(\ref{deltal}) in the form
 \[\label{deltallz}
 {\Delta}_{n,l}(t)={ {\ts {Z^{l}-Z^{-l}}\over{\ts Z^{1}-Z^{-1}}}
 {{ \ts t^{1\over 2}-t^{-{1\over 2}}}\over{\ts
t^{l\over2}-t^{-{l\over 2}}}}}= \]
\[
= {\cal F}\bigl(Z(t,\Delta_{n,2}(t)~); l, t\bigr) = {\cal
F}\bigl(\Delta_{n,2}(t); l,t\bigr)
 \,,
 \]
which means that all the functional dependence on $n$ in the
Alexander polynomial $\Delta_{n,l}(t)$ is contained in
$\Delta_{n,2}(t)\,$ alone, $n$ being coprime with $l$ and $2$. For
example, the whole dependence on $n$ for $\Delta_{7,5}(t)$ is in
$\Delta_{7,2}(t)$, but in $\Delta_{8,3}(t)$ for $\Delta_{8,5}(t)$
(for the latter type connection, see below).

Yet another formula, which means that the explicit dependence on $n$
for the Alexander polynomial $\Delta_{n,l}(t),$ is given by
$\Delta_{n,k}(t)\,,\,k=2$ or $3$, with respective $n$, follows from
combining (\ref{dnl-v3}), (\ref{deltal2}), and (\ref{deltal3}),
 namely:
 \be\label{dnl-v3--}
  \Delta_{n,l}(t)={\frac{V_{l-1}(z)}{V_{l-1}(y)}},
\qquad y{=}t^{1\over2}+t^{-{1\over2}},
 \ee
where if $k=2\,$ and $n$ is coprime with $l$ and $2$,  we have that
 \[
 z=(t^{\frac12}+t^{-\frac{1}{2}})\Delta_{n,2}(t)\,.
  \]
At last, if $k=3\,$ and $n$ is coprime with $l$ and $3$, then we have, in
(\ref{dnl-v3--}), that
 \[
 z=\Bigl((t+1+t^{-1})\Delta_{n,3}(t)+1\Bigr)^{\frac12}\,.
  \]

\section{Concluding Remarks}

In this paper, we have shown how to express any Alexander polynomial
$\Delta_{n,3}(t)$ as a (finite) sum of the summands
 $\Delta_{k,2}(t)\,$ with appropriate values of $k$,
 see formulas (\ref{t2}) and (\ref{exam32}).
 Note that if one succeeds to solve the similar problem of
 expressing $\Delta_{n,l}(t)$ by the sum of $\Delta_{k,2}(t)\,$,
 it would give us some novel, interesting description for a general torus knot.

We have demonstrated for this, more general case of torus knots than
the one treated in \cite{GP}, the close connection of the Alexander
polynomials $\Delta_{n,l}(t)$ with the Chebyshev polynomials.
  Among the others, we recall the result given by (\ref{dnl-v2-})
   showing that the dependence on $n$ in the $\Delta_{n,l}(t)$
 can be incorporated just in the Chebyshev polynomial of the second kind,
 whereas the dependence on $l$ -- in the Chebyshev polynomial of the first kind,
 the latter being the argument of the former one.  
  We have also shown that the dependence of $\Delta_{n,l}(t)$ on $n$
 may be through $\!\Delta_{n,2}(t)$\ ($n, l, 2$ are coprime) or
 through $\!\Delta_{n,3}(t)$  ($n, l, 3$ are coprime).

In the visible future, we hope to find a particular physical
realization of the obtained results in the direction of constructing
the knot-like configuration(s) within some field theory model.

\vskip3mm
 This research was partially supported by the Grant 29.1/028 of
 the State Foundation of Fundamental Research of Ukraine
and by the Special Program of the Division of Physics and Astronomy
of NAS of Ukraine.

\rezume{%
ПОЛІНОМІАЛЬНІ ІНВАРІАНТИ АЛЕКСАНДЕРА \\ТОРИЧНИХ ВУЗЛІВ $T(n,3)$ І
ПОЛІНОМИ ЧЕБИШОВА}
{%
О.М. Гаврилик, А.М. Павлюк} {Отримана явна формула, яка виражає
поліноми Александера $\Delta_{n,3}(t)$ торичних вузлів $T(n,3)$
через суму поліномів Александера $\Delta_{n,2}(t)$ торичних вузлів
$T(n,2)$.
  На основі цього, а також результатів наших попередніх  робіт,
  ми виражаємо поліноми Александера $\Delta_{n,3}(t)$ в термінах
  поліномів Чебишова. Даний результат поширено на довільні торичні
 вузли $T(n,l)$, де $n$ та $l$ -- взаємно прості цілі числа.}

\end{document}